\documentclass[runningheads]{llncs}

\usepackage[american]{babel}
\usepackage{cite}
\usepackage{amsmath,amssymb,amsfonts}
\usepackage{algorithmic}
\usepackage{graphicx}
\usepackage{textcomp}
\usepackage{xcolor}
\usepackage[utf8]{inputenc}
\usepackage[T1]{fontenc}
\usepackage{listings}
\usepackage{url}
\usepackage{array}
\usepackage{booktabs}
\usepackage{subcaption}
\def\BibTeX{{\rm B\kern-.05em{\sc i\kern-.025em b}\kern-.08em
    T\kern-.1667em\lower.7ex\hbox{E}\kern-.125emX}}

\usepackage{multirow}
\usepackage{graphicx}
\usepackage{cleveref}
\usepackage{enumitem}

\newcolumntype{M}[1]{>{\centering\arraybackslash}p{#1}}

\usepackage{draftwatermark}
\SetWatermarkText{\bfseries PREPRINT}
\SetWatermarkScale{3.5}
\SetWatermarkLightness{0.9}

\begin{document}

\title{Architecture design of a networked music performance platform for a chamber choir}
\titlerunning{Architecture design of a networked music performance platform...}

\author{
Jan Cychnerski\orcidID{0000-0003-2733-4599},
Bartłomiej Mróz\orcidID{0000-0002-8983-9050}
}

\authorrunning{J. Cychnerski, B. Mróz}

\institute{Faculty of Electronics, Telecommunications and Informatics \\
Gdańsk University of Technology, Poland \\
jan.cychnerski@eti.pg.edu.pl, bartlomiej.mroz@pg.edu.pl}

\maketitle

\vspace{-5mm}

\begin{abstract}

This paper describes an architecture design process for Networked Music Performance (NMP) platform for medium-sized conducted music ensembles, based on remote rehearsals of Academic Choir of Gdańsk University of Technology. The issues of real-time remote communication, in-person music performance, and NMP are described. Three iterative steps defining and extending the architecture of the NMP platform with additional features to enhance its utility in remote rehearsals are presented. The first iteration uses a regular video conferencing platform, the second iteration uses dedicated NMP devices and tools, and the third iteration adds video transmission and utilizes professional low-latency audio and video workstations. For each iteration, the platform architecture is defined and deployed with simultaneous usability tests. Its strengths and weaknesses are identified through qualitative and quantitative measurements -- statistical analysis shows a significant improvement in rehearsal quality after each iteration. The final optimal architecture is described and concluded with guidelines for creating NMP systems for said music ensembles.

\end{abstract}

\begin{keywords}
real-time system, networked music performance, platform design

\end{keywords}

\section{Introduction and related work}

In recent years, advances in technology have resulted in the increasing use of real-time remote working devices. Also, the COVID-19 pandemic made many aspects of everyday life transfer to the virtual world. For some fields of work this change does not affect the efficiency and quality of work. Nevertheless, remote creative and artistic work, particularly music performance, remains a challenge.

\label{sec:intro}

\subsection{Real-time communication}

Recently, various real-time conferencing tools (e.g. Zoom, MS Teams, Google Meet) rapidly expanded and amply gained new users \cite{matulin2021comparison}. The success of conferencing platforms was largely caused by the general features that make typical Internet communication (Fig. 1) simple and effective, which include \cite{correia2020evaluating,faadhilah2022comparison}: ease of installation, convenience of use, ability too operate under varying network conditions, resistance to audio noise (low quality microphone, background noises), video and screen sharing, chat, file transfer, shared whiteboard, integration with office/school software.

In order to achieve the aforementioned goals, these applications must exhibit very high automation and compatibility, which is achieved at a cost of \cite{hossfeld2008analysis}:

\begin{enumerate}
    \item long buffers, high transmission delays reaching several hundred milliseconds
    \item very aggressive noise reduction, cutting out a large part of the bandwidth, and even changing the audio signal content using AI
    \item strong, automated emphasis on the main speaker while muting the others
\end{enumerate}

\begin{figure}[htbp]
    \centering
    \includegraphics[width=0.6\linewidth]{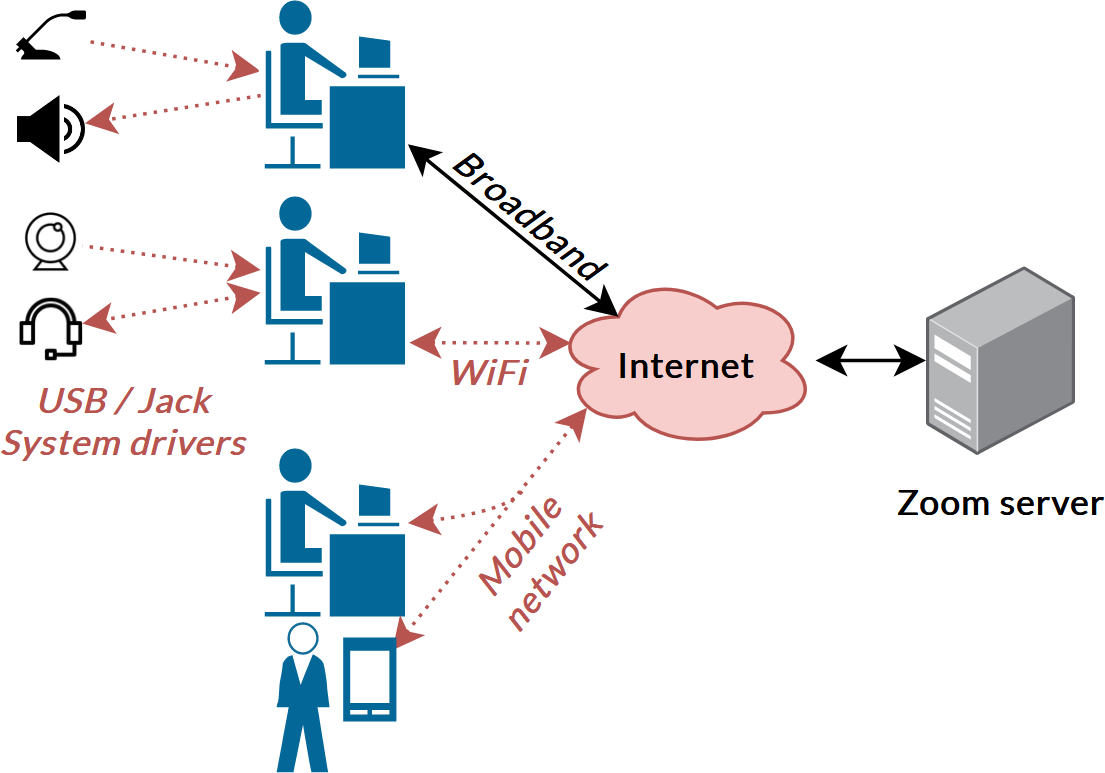}
    \caption{Typical real-time communication platform architecture}
    \label{fig:realtime-communication}
\end{figure}

\subsection{In-person music performance}

A different set of requirements is posed by music performance in typical medium-sized ensembles (e.g. chamber choirs), especially those with a conductor. During normal stationary in-person rehearsals, communication is characterized by \cite{king2004collaboration,pennill2019ensembles,volpe2016measuring}:

\begin{enumerate}
    \item no audiovisual delays of any kind -- the performers are in the same room 
    \item the performers hear each other and adjust their performance in real time to the close performers and the overall sound of the ensemble (self-feedback)
    \item a conductor in front of the performers ensures overall cohesion, interpretation, volume and tempo, conducting visually in real time
    \item the physical location of the performers matters -- people standing closer (so performing similar musical parts) are heard louder (intra-section feedback)
\end{enumerate}

\begin{figure}[htb]
    \centering
    \includegraphics[width=0.4\linewidth]{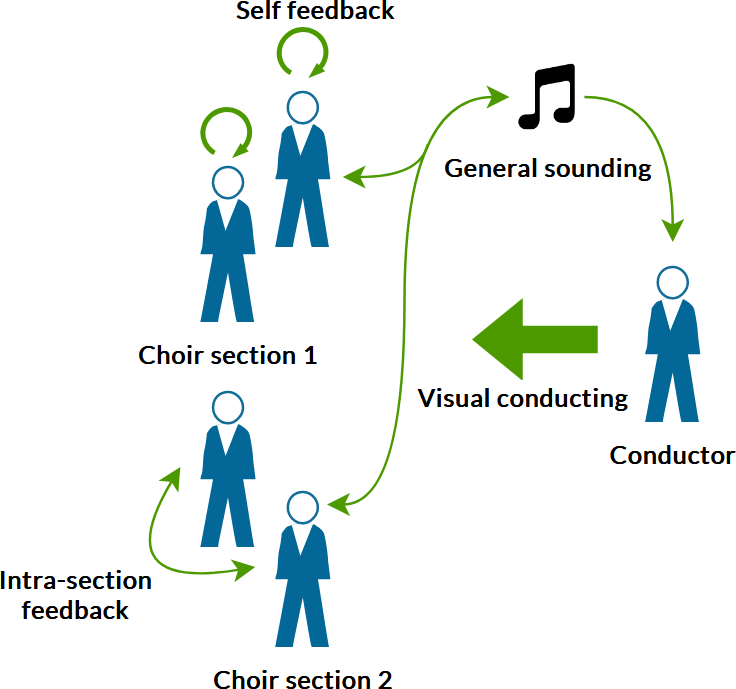}
    \caption{Typical stationary in-person music performance}
    \label{fig:music-performance}
\end{figure}

\subsection{Networked music performance}

The features of typical conferencing tools are highly inadequate for a music performance. For this reason, the topic of remote music performance is a separate branch in research and industry, called Networked Music Performance \cite{Dessen2020nmpintro}. In recent years, advances in technology have resulted in the increasing use of real-time remote working devices. Nevertheless, remote creative and artistic work, particularly music performance, remains a challenge. One-way connections, through which artistic events are transmitted in real-time, have become available using networks capable of supporting high bandwidth, low-latency packet routing, and guaranteed quality of service (QoS) \cite{chafe2000simplified,chafe2009tapping}. These can be described as unidirectional, allowing artists to remotely connect only between the performance venue and the audience, not each other. It is much more difficult when performers in two or more remote locations attempt to perform an established composition or improvisation together in real-time. Weinberg \cite{weinberg2005interconnected} calls this way of performing music the “bridge approach”. In such cases, the inevitable delay caused by the physical transit time of network packets is known to affect performance \cite{bartlette2006effect,bouillot2009challenges,chafe2010effect,gu2005network}. Therefore, attempts have been made to account for these delays by design or composition \cite{bouillot2007njam,caceres2008edge}.

A number of NMP experiments and frameworks are described in \cite{rottondi2016overview,mall2021low,onderdijk2021impact}. Some platforms, like MusiNet or Diamouses, are large-scale projects run by universities; other, like LoLa or JackTrip require a subscription or licensing plan. The Jamulus platform stands out as an open-source, community-driven and easy to use NMP tool \cite{fischercase}. It allows joint performance via freely available servers, or via dedicated server configured with Jamulus software on a local machine. These aspects are especially appealing to smaller, non-professional ensembles (e.g. students, local communities, etc.) which are not capable of huge financial investments or involvement in research projects regarding remote rehearsing.

\subsection{Contribution}

This paper discusses the process of defining and practically realizing a platform architecture that meets the requirements for networked music performance of medium-sized conducted amateur music ensembles through the example of remote rehearsals of a chamber choir. The proposed architecture in contrast to existing publications and commercial solutions takes into account: (1) feasibility study in a highly ecologically valid, choral environment (vs artificial environments in NMP papers \cite{rottondi2016overview}), (2) use of available, open-source platforms for low-latency transmission of both audio and video (vs difficult to obtain, scientific projects like Internet2, Musinet, Diamouses, GigaPoP, etc.), (3) consideration for the lack of technical capabilities of performers, (4) consideration for a low-cost hardware options, (5) proposal for an architecture for the NMP and a set of good practices.

\section{Experimental setup}

In the experiment, 3 NMP architectures were proposed and investigated for a $\sim$20-member choral ensemble: Zoom@Home, Jamulus@Home and Jamulus@Univ. Each architecture was implemented in practice as means for the rehearsals of the Academic Choir of the Gdansk University of Technology, during the 2021 COVID-19 outbreak in Poland. After deployment, the architectures were used without major modifications for nearly 2 months each. Architectures were evaluated qualitatively by the choristers through a survey and by the authors responsible for the implementation. Quantitative assessment was performed through a round trip time (RTT) audio and video latency measurements. After each deployment, various aspects affecting rehearsal quality were determined, and based on them, decisions were made to implement modifications in the next iteration. Comparative features of all architectures are shown in Table \ref{tab:architectures}, and described in detail in the following subsections.

\begin{table}[htb]
\caption{Comparison of proposed architecture features}
\label{tab:architectures}
\footnotesize
\begin{tabular}{M{3cm}M{3cm}M{3cm}M{3cm}}
\toprule
\textbf{Architecture}           & \textbf{Zoom@Home}                                                       & \textbf{Jamulus@Home}                                                                    & \textbf{Jamulus@Univ}                   \\ \toprule
\textbf{Connection}             & any                                                                 & broadband                                                                                & LAN                                     \\ \midrule
\textbf{Choir audio}     & any                                                                 & Semi-professional; ASIO drivers                                        & professional; ASIO interface      \\ \midrule
\textbf{Conductor audio} & any                                                                 & \multicolumn{2}{M{6cm}}{professional, using ASIO interface}                                                                             \\ \midrule
\textbf{Audio RTT}          & 300-1000 ms                                                         & 63-135 ms                                                                                & 40-85 ms                                \\ \midrule
\textbf{Choir video}     & any                                                                 & --                                                                                        & --                                       \\ \midrule
\textbf{Conductor video} & any                                                                 & --                                                                                        & low latency camera                      \\ \midrule
\textbf{Video latency}          & 500-1000 ms                                                         & --                                                                                        & 25-100 ms                              \\ \midrule
\textbf{Feedback}               & --                                                                   & \multicolumn{2}{M{6cm}}{self \& intra-section feedback, choir mix}                                                      \\ \midrule
\textbf{Setup}                  & automatic, self-supervised                                          & manual                                                                                   & manual, semi-supervised                 \\ \midrule
\textbf{Assistance}             & --                                                                   & remote                                                                                   & remote+local                   \\ \midrule
\textbf{Rehearsal scope}             & individual singing with muted mic, solo singing & singing section parts or tutti with piano & Jamulus@Home + singing tutti a'cappella \\ \bottomrule
\end{tabular}
\end{table}

\subsection{Zoom@Home architecture: a simple conferencing tool}

The first proposed architecture for remote choir rehearsals was a standard web-based communication platform using popular tools. Therefore, Skype, Jitsi, MS Teams and Zoom platforms were initially tested. Eventually Zoom was selected due to its popularity, ease of use, compatibility and very low entry threshold. No changes were made to the typical conferencing architecture, as shown in Fig. \ref{fig:realtime-communication}. 
Simultaneous singing was effectively impossible due to significant delays and the automated speaker highlight feature that amplifies one person and mutes the others. The choir could rehearse only in the following capacities: (1) casual conversation as in a typical conference, where everyone has microphone on and everyone can talk, (2)  singing together with microphones turned off; the only active microphone is the conductor's, who leads the rehearsal by playing choral parts on the piano, and (3) solo singing with occasional help and commentary from the conductor.
From a technical standpoint, the Zoom@Home architecture posed no problems. There were no requirements for hardware, Internet connection, or location. Installation and configuration were possible on any device; each chorister configured the Zoom application without technical assistance; most choristers used a webcam to enhance the feeling of presence at a rehearsal. RTT latencies ranged from 300-1000 ms for the audio channel and 500-1000 ms for the video channel. Such high RTTs prevented more than one person from singing at a time, making it impossible to measure relative latency between choristers.

\subsection{Jamulus@Home architecture: NMP with choristers at home}

\begin{figure}[h]
    \centering
    \includegraphics[width=0.75\linewidth]{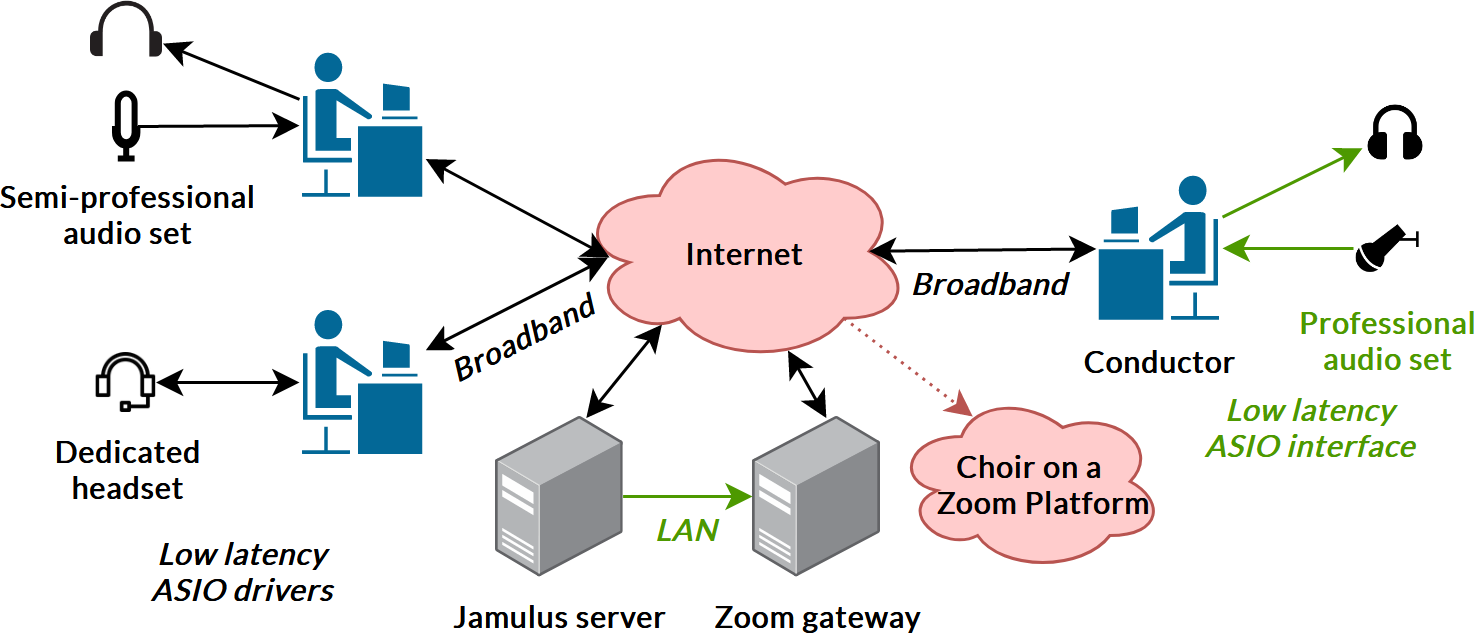}
    \caption{Jamulus@Home NMP platform architecture}
    \label{fig:jamulus-home}
\end{figure}

The second iteration of the architecture -- presented in Fig. \ref{fig:jamulus-home} -- introduced several improvements. Jamulus software was chosen due to its main features: open-source, cross-platform, built by the NMP community, focused on low latency audio connection, no limitations on the number of participants, and no requirements for specialized hardware. Several inexpensive headsets were reviewed for latency, compatibility, sound quality, and comfort; then one optimal model (Logitech PC 960 USB) was selected. Additionally, semi-professional low-latency compatible audio equipment was allowed. To ensure minimal delays and high quality of the conductor's audio connection, he was provided with professional audio equipment. The video connection got discarded, further lowering the latency; a dedicated Jamulus server was deployed at the University, ensuring broadband connection in the nearest area. A requirement was imposed for a minimum internet connection type (wired broadband) and a maximum geographical distance from the Jamulus server (100 km). A correctly configured low-latency ASIO driver was required. That resulted in RTT in the range of 63-135 ms, which allowed for real-time collaborative singing.

Despite attempts to unify the hardware and imposed requirements, the hardware and software configuration had to be done manually by users, with active remote support (via Windows Remote Assitance or TeamViewer) provided by two designated technicians. The real-time NMP solution required a complete revision of remote workflow. Simultaneous singing became viable -- it allowed the whole ensemble to sing together (tutti) for the first time under pandemic restrictions. Mechanisms included in the Jamulus software enabled the choristers to adjust the volume of particular persons, allowing them for intra-section feedback (strengthening their sections and weakening the rest); similar to traditional rehearsals. The conductor could communicate with the choristers on a real-time basis and lead through accompaniment on a piano. It became possible to rehearse in sections and hear the overall sound of the pieces, allowing meaningful work on the repertoire

Not all choristers could meet the requirements of NMP (equipment, connectivity, etc.). A solution to this problem was an additional gateway server to the Zoom platform. As a countermeasure for high delays in the audio signal of the Zoom platform, the communication with the gateway server was one-way. Thus, it allowed choristers not capable of NMP for passive participation in rehearsals.

\subsection{Jamulus@University architecture: low-latency audio+video, choristers and conductor at the University}


\begin{figure}[htbp]
    \centering
    \includegraphics[width=0.75\linewidth]{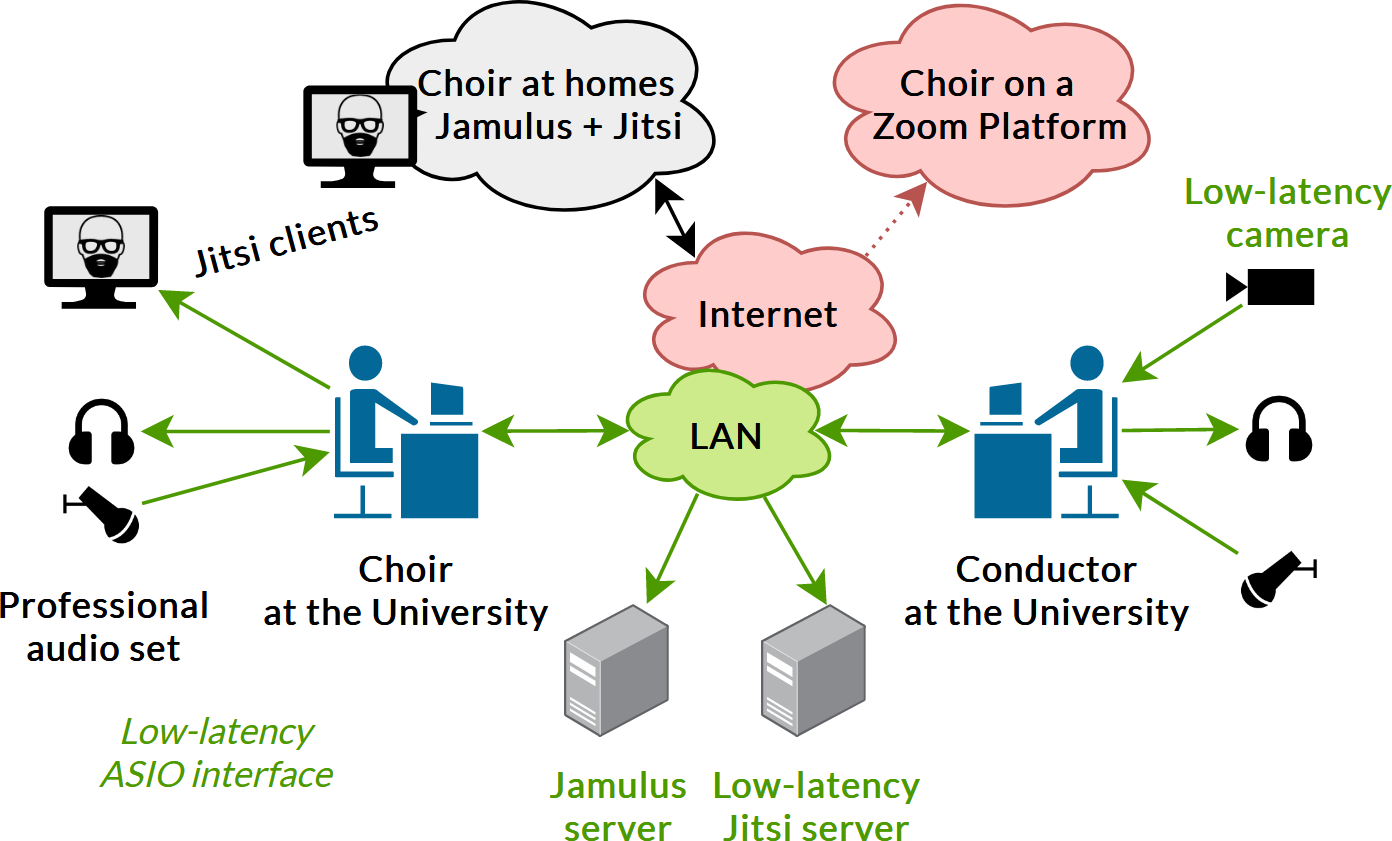}
    \caption{Jamulus@University NMP platform architecture}
    \label{fig:4}
\end{figure}


Despite the ability to sing together in the previous iteration, persisting transmission delays caused frequent loss of synchronization between choristers, requiring the conductor to enforce the tempo through either accompaniment or counting aloud. Still, maintaining a steady performance was very difficult and unattainable without a conductor. Moreover, due to the lack of video input, it was impossible for the conductor to lead visually, making it impossible for him to control the tempo variability of the pieces (accelerations, decelerations, fermatas, tempo rubato).

In the third iteration of the architecture design, namely Jamulus@University, choristers got invited to a dedicated classroom at the university, equipped with pre-configured professional audio hardware, providing high quality audio with very low latency. The conductor's workstation was supplied with low-latency audio-video equipment and moved to another classroom with a piano. In order to ensure a low latency video connection, a dedicated Jitsi server was deployed, providing the conductor's video with a latency of 25-100 ms. Such a delay allowed the conductor to feasibly lead the choir in a real-time manner visually, which was previously inaccessible. All communication was done over a LAN, reducing the audio latency to 40-85 ms RTT. Controlled environment allowed creation of semi-automated scripts facilitating deployment of necessary software for workstations (2-person assistance was still in place, but in smaller capacity).

The Jamulus@University was the most extensive of the architectures tested, as presented in Fig \ref{fig:4}. It was used for two months of remote rehearsals and was crucial in sustaining choral activities during the COVID-19 lockdown.

\section{Results}

Each iteration contained a questionnaire assessing NMP architecture's usability. Since the remote rehearsals lasted several months, some choristers participated in several iterations. In total, 23 choir members (9 male, 14 female, aged 18-30) participated in the experiments. The Zoom@Home architecture assessment consists of 8 responses; the Jamulus@Home and the Jamulus@University assessments comprise 16 and 13 answers, respectively. All questions and answer scales are presented in \ref{tab:questions}.

\begin{table}[htbp]
\centering
\footnotesize
\caption{Questions and answer scales in the questionnaire}
\label{tab:questions}
\begin{tabular}{ccc}
\toprule
\textbf{Question}     & \textbf{Applies to}                                                                                    & \textbf{Likert scale description}                                                                                                  \\ \midrule
Excersiing difficulty & \multirow{2}{*}{\begin{tabular}[c]{@{}c@{}}Zoom@Home\\ Jamulus@Home\\ Jamulus@University\end{tabular}} & \begin{tabular}[c]{@{}c@{}}1-3 - more difficult\\ 4 - about the same\\ 5-7 - easier\end{tabular}                                   \\ \midrule
Rehearsing comfort    &                                                                                                        & \begin{tabular}[c]{@{}c@{}}1-3 - less comfortably\\ 4 - about the same\\ 5-7 - more comfortably\end{tabular}                       \\  \midrule
Setup difficulty      & Jamulus@Home                                                                                           & \begin{tabular}[c]{@{}c@{}}1 - very difficult, a lot of help needed\\ 5 - very easy, everything highly understandable\end{tabular} \\
Rehearsal safety      & Jamulus@University                                                                                     & \begin{tabular}[c]{@{}c@{}}1 - strong fear of infection\\ 5 - no fear of infection\end{tabular}                                    \\  \bottomrule
\end{tabular}
\end{table}

\subsection{Assessment of the Zoom@Home architecture}

\vspace{-2mm}
\begin{figure}[htbp]
	\centering
	\begin{subfigure}[b]{0.30\textwidth}
	    \includegraphics[width=\textwidth]{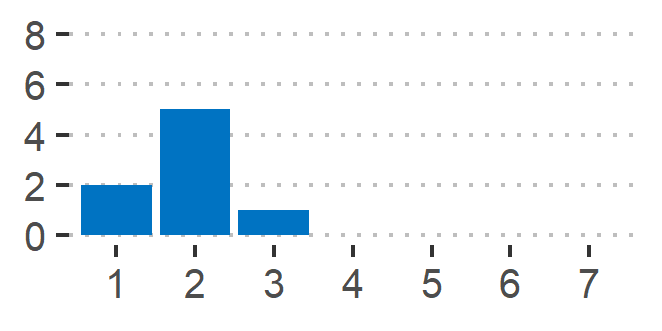}
	    \caption{}
	\end{subfigure}
	\begin{subfigure}[b]{0.30\textwidth}
	    \includegraphics[width=\textwidth]{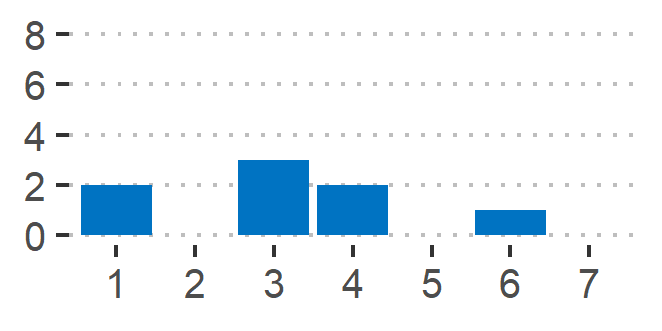}
	    \caption{}
	\end{subfigure}
	\caption{Exercising difficulty and rehearsing comfort of the Zoom@Home NMP, respectively. Likert scale: 1-3 = more difficult / less comfortable, 4 = about the same, 5-7 = easier / more comfortable}
	\label{fig:results-zoom-home}
\end{figure}
\vspace{-2mm}

The first examined was the Zoom@Home NMP architecture. As it did not differ much from typical conferencing platforms, only two rehearsal quality aspects were measured -- exercising difficulty and rehearsing comfort, referring to traditional, non-remote rehearsals -- as shown in Fig. \ref{fig:results-zoom-home}. Most of the participants rated the difficulty as 2: more difficult than on traditional rehearsal. Some participants reported shyness with solo singing; the rating might reflect that intricacy. Interestingly, the rehearsing comfort was rated mainly as 3: a bit less comfortably, than on traditional rehearsal, and 4: about the same, as on traditional rehearsal. Many choristers reported high comfort in rehearsing at home, especially those with conducive conditions. However, some participants reported difficulties when singing at home; the most highlighted were other household members, neighbors getting easily irritated, or a noisy environment with construction or traffic noise from nearby places.

\subsection{Assessment of the Jamulus@Home architecture}


\begin{figure}[htbp]
	\centering
	\begin{subfigure}[b]{0.30\textwidth}
	    \includegraphics[width=\textwidth,height=2cm]{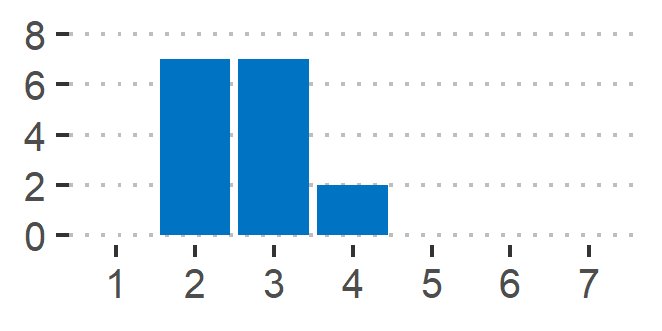}
    	\caption{}
    \end{subfigure}
    \begin{subfigure}[b]{0.30\textwidth}
	    \includegraphics[width=\textwidth,height=2cm]{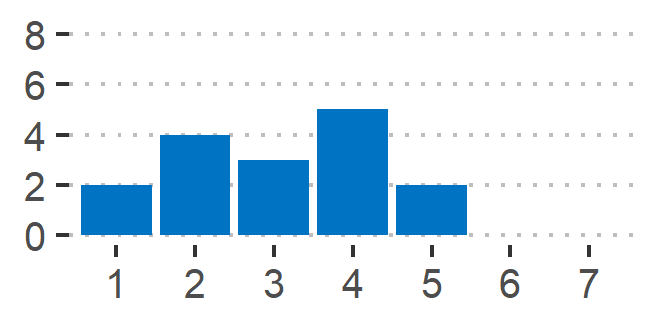}
    	\caption{}
    \end{subfigure}
    \begin{subfigure}[b]{0.30\textwidth}
        \includegraphics[width=\textwidth,height=2cm]{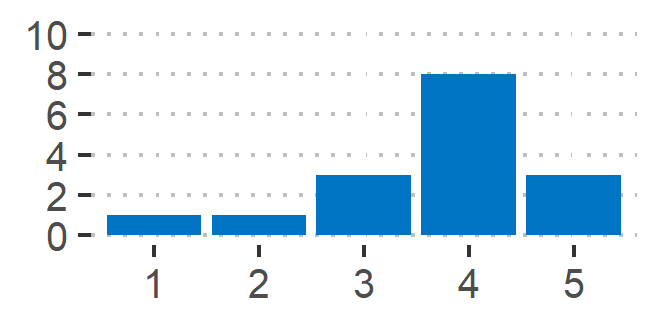}
    	\caption{}
    	\label{fig:results-setup}
    \end{subfigure}
    \caption{Exercising difficulty, rehearsing comfort, and setup difficulty of Jamulus@Home NMP. Likert scale for (c): 1 =  very difficult, a lot of help needed; 5 = very easy, everything highly understandable.}
    \label{fig:results-jamulus-home}
    
\end{figure}


The Jamulus@Home architecture was evaluated by the following quantitative factors: exercising difficulty and rehearsing comfort as compared to traditional, non-remote rehearsals, inter-chorister latency, and setup difficulty. Most participants rated the difficulty as 2: more difficult, and 3: a bit more difficult, than on traditional rehearsals. This rating appears slightly higher than for the previous platform. The ratings for rehearsing comfort are somewhat higher than for the Zoom platform; most choristers rated the comfort as 4: about the same, as on traditional rehearsal. This architecture allowed inter-chorister latency measurement: the time difference of singing the same note was 83 ms \textpm 57 ms. Furthermore, the setup difficulty rating concentrated around 4 -- as shown in Fig. \ref{fig:results-setup} -- indicating minor setup difficulties with remote assistance required.

\subsection{Assessment of the Jamulus@University architecture}

\vspace{-4mm}

\begin{figure}[htbp]
    \centering
    \begin{subfigure}[b]{0.3\textwidth}
    	\centering
    	\includegraphics[width=\textwidth,height=2cm]{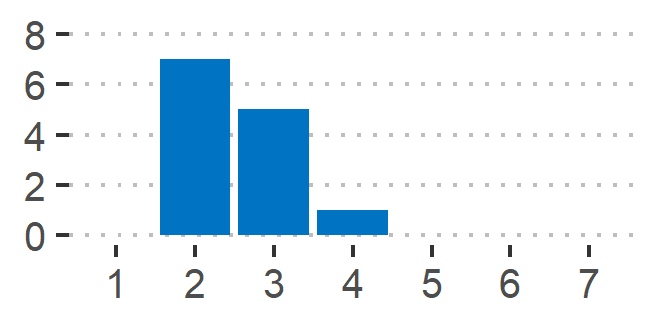}
    	\caption{}
    \end{subfigure}
    \begin{subfigure}[b]{0.3\textwidth}
    	\includegraphics[width=\textwidth,height=2cm]{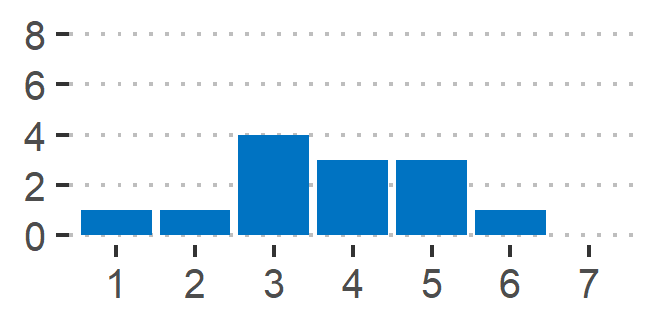}
    	\caption{}
    \end{subfigure}
    \begin{subfigure}[b]{0.3\textwidth}
    	\includegraphics[width=\textwidth,height=2cm]{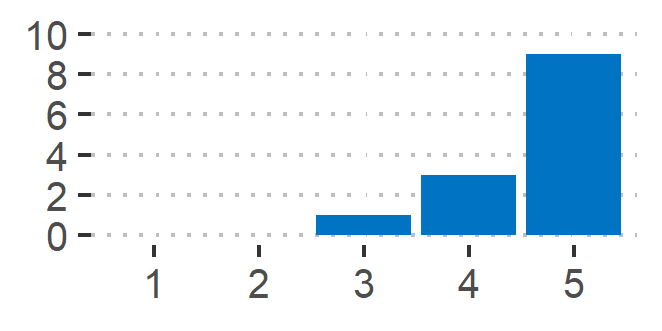}
    	\caption{}
    	\label{fig:results-safety}
    \end{subfigure}
	\caption{Exercising difficulty, rehearsing comfort and safety of Jamulus@University. Likert scale for (3): 1 = strong fear of infection, 5 = no fear of infection}
	\label{fig:results-jamulus-univ}
\end{figure}

\vspace{-3mm}

The Jamulus@University assessment comprised exercising difficulty and rehearsing comfort ratings in relation to traditional, non-remote rehearsals (Fig. \ref{fig:results-jamulus-univ}), as well as inter-chorister latency. Most participants rated the difficulty as 2: more difficult, and 3: a bit more difficult, than on traditional rehearsals. The ratings for rehearsing comfort seem to be more spread but mainly concentrated around score 4: about the same, as on traditional rehearsal.

Higher comfort resulted also from a reduced sense of shyness in these conditions. Additionally, the choristers were asked for a rating of perceived safety, since it involved spending a considerable amount of time in one classroom. The majority of the choristers rated the safety as 5: no fear of infection (Fig. \ref{fig:results-safety}). Such a result might reflect the convenient placement of dedicated workstations in separate compartments made with plywood walls. This setup gave the choristers a level of separation, especially in terms of loudness and personal distance. It is also worth noting that no COVID-19 outbreak occurred within the choir at that time.

With reduced latency and the addition of video, it became possible to perform tutti a'cappella pieces without accompaniment, metronome, or other enforced tempo control. Overall, keeping the tempo became easier, even though some of the choisters were still using the Jamulus@Home platform. The time difference between different choristers singing the same note dropped to 47 \textpm 46 ms.

\subsection{Statistical analysis}

\begin{figure}[htbp]
	\centering
	\begin{subfigure}[b]{0.40\textwidth}
	    \includegraphics[width=\textwidth]{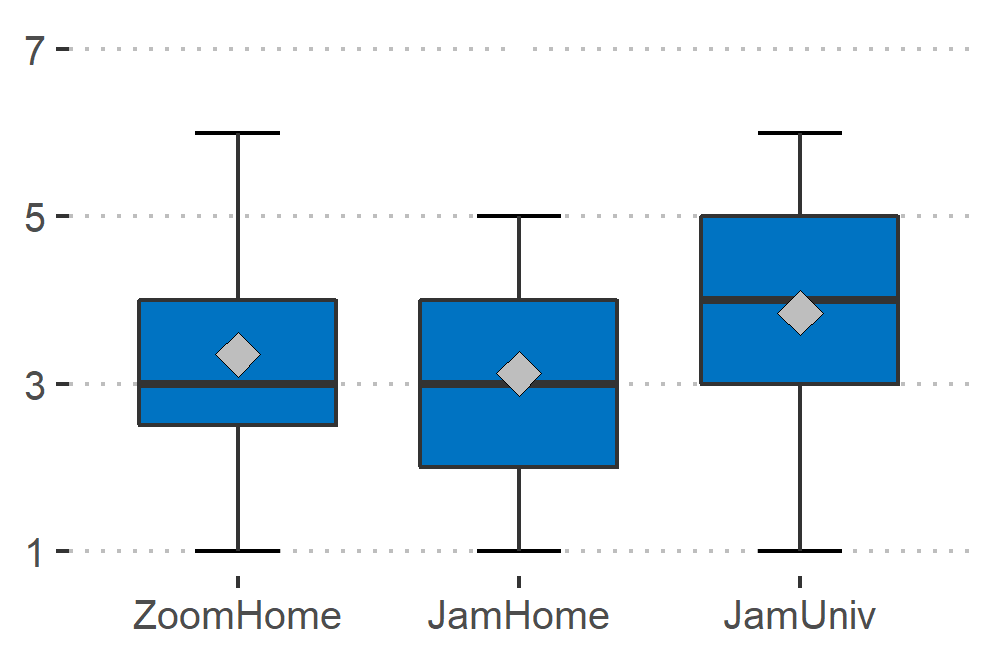}
	    \caption{}
	\end{subfigure}
	\begin{subfigure}[b]{0.40\textwidth}
	    \includegraphics[width=\textwidth]{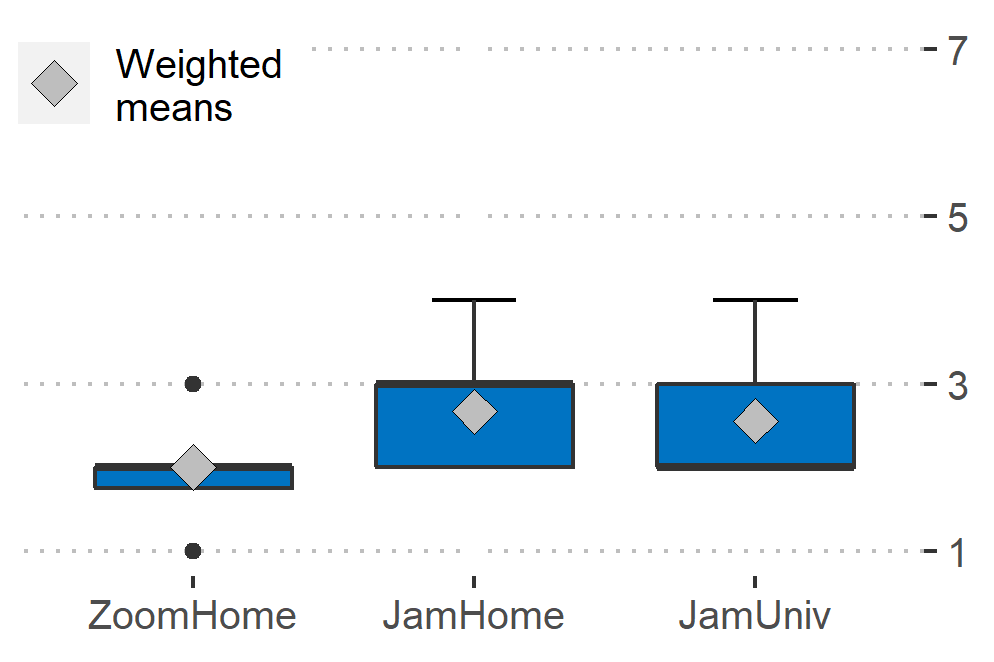}
	    \caption{}
	\end{subfigure}
	\caption{Rehearsing comfort and exercising difficulty statistics}
	\label{fig:results-stats}
\end{figure}

All statistical calculations were performed with R software (version 4.1.2). The most notable packages were MASS (7.3-54), car (3.0-12), emmeans (1.7.1-1), ordinal (2019.12-10) and RVAideMemoire (0.9-80).
The assessment of exercising difficulty and rehearsing comfort was repeated for each architecture's iteration, allowing direct comparison. Some participants answered more than one question; therefore, a generalized linear mixed model approach was selected. Since the answers were on a 7-point Likert scale, the Cumulative Link Mixed Model (CLMM) was chosen for the analysis. 
The dependent variable, Platform, being a categorical variable, was encoded using Helmert contrast coding. Also, the analysis was performed on both unweighted and weighted response data. The weights were derived from the number of participants and were assigned as follows: 1-2 rehearsals -- 1; 3-6 rehearsals -- 2; 7 and more rehearsals -- 3. Results are shown in Fig. \ref{fig:results-stats}.

\vspace{-3mm}

\vspace{-3mm}

An analysis of variance for unweighted data based on ordinal logistic regression indicated no statistical effect on rehearsing comfort ($\chi^2$(2,N=37) = 1.93, n.s.) However, testing on weighted data indicated a statistical effect ($\chi^2$(2,N=86) = 10.84, $p$ < 0.01). Since the statistical significance was detected on weighted data, post-hoc analysis was performed for this scenario. Pairwise comparisons using Z-tests, corrected with Holm’s sequential Bonferroni procedure, indicated that Likert scores for Jamulus@Home vs. Jamulus@Univ were statistically significantly different (Z = -3.1, $p$ < 0.01), but Likert scores for Jamulus@Home vs. Zoom@Home and for Jamulus@Univ vs. Zoom@Home were not significantly different (Z = -1.4, n.s., and Z = 1.7, n.s.).

The statistical analysis is consistent with what is shown in Fig. \ref{fig:results-stats}a – the rehearsing comfort is similar for participants in all three platforms. However, weighted Likert scores induced a statistically significant difference between Jamulus@Home platform and Jamulus@Univ platform, suggesting higher comfort of rehearsing for the latter.

An analysis of variance for unweighted data based on ordinal logistic regression indicated a statistical effect on exercising difficulty ($\chi^2$(2, N=37) = 29.5, $p$ < 0.001). For weighted data, this test also indicated a statistical effect ($\chi^2$(2, N=86) = 46.4, $p$ < 0.001). Thus, post-hoc analysis was performed for both cases. For unweighted Likert scores, pairwise comparisons using Z-tests, corrected with Holm’s sequential Bonferroni procedure, indicated that Likert scores for Jamulus@Home vs. Jamulus@Univ were statistically significantly different (Z = -1910, $p$ < 0.001), also Likert scores for Jamulus@Home vs. Zoom@Home and for Jamulus@Univ vs. Zoom@Home were statistically significantly different (Z = 24025, $p$ < 0.001, and Z = 29152, $p$ < 0.001). The comparisons for weighted Likert scores indicated the higher magnitude of statistically significant differences: Jamulus@Home vs. Jamulus@Univ (Z = -6494, $p$ < 0.001), Jamulus@Home vs. Zoom@Home Z = 31852, $p$ < 0.001), and Jamulus@Univ vs. Zoom@Home (Z = 42875, $p$ < 0.001).
The statistical analysis is somewhat inconsistent with what is in Fig. \ref{fig:results-stats}b -– the exercising difficulty on the Zoom@Home platform is clearly higher than on any of the Jamulus platforms, but both Jamulus platforms seem to have the same exercising difficulty. However, the statistical analysis showed for both weighted and unweighted Likert data a significant difference between them, in favor of the Jamulus@Univ one; adding weights increased the magnitude of this difference. That would be consistent with choristers’ opinions about this platform.

\subsection{Proposed architecture, recommendations and good practices}
\label{sec:proposed-architecture}

\begin{figure}[htbp]
	\centering
	\includegraphics[width=0.7\textwidth]{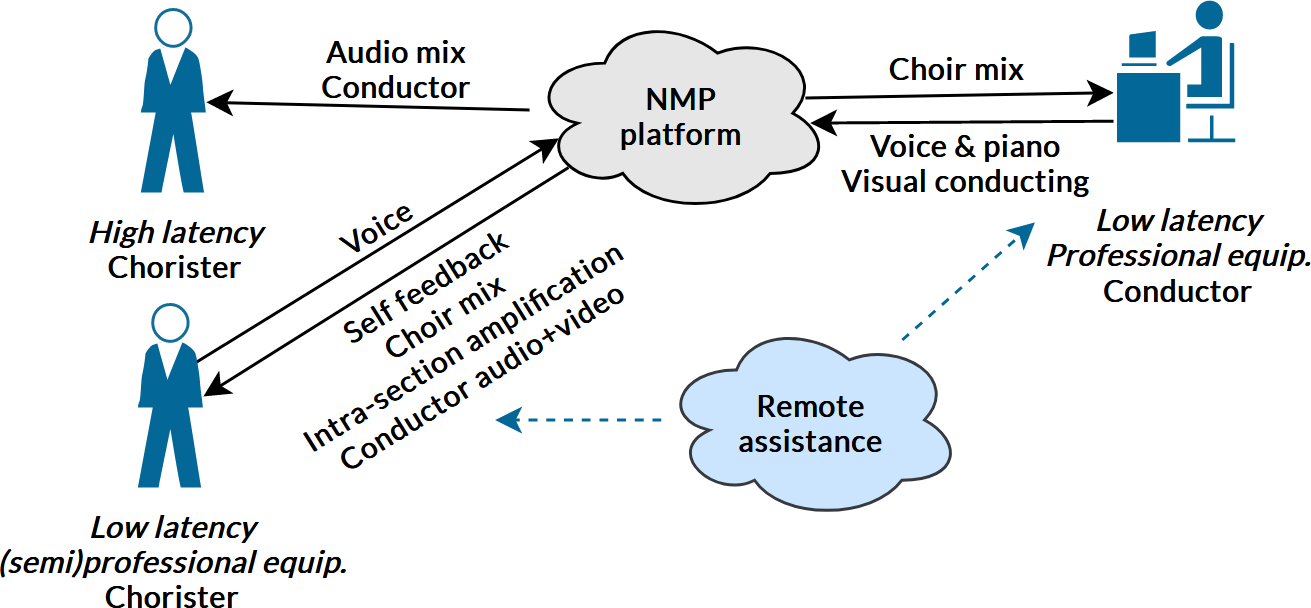}
	\caption{Proposed generalized NMP platform architecture}
	\label{fig:proposed-architecture}
\end{figure}

The experiments and analysis of the deployed architectures' assessments concluded with the construction of the optimal architecture of the NMP platform for chamber ensembles with a conductor. The general architecture should fulfill the following requirements and good practices (also shown in Fig. \ref{fig:proposed-architecture}):

\begin{enumerate}
    \item affordable, cross-platform, low-latency (RTT not exceeding 100 ms) NMP software, along with a dedicated server located near the performers;
    \item dedicated, low-latency (not exceeding 100 ms) video streaming server;
    \item conductor's station with professional equipment, fast network, close to the servers, allowing unrestricted spoken, accompanied and visual conducting;
    \item performer stations with professional audio equipment, located close to the servers, allowing for adjustable self-feedback, the ability to hear the overall ensemble's mix, with additional amplification of their own section;
    \item performers who cannot use (4) should be equipped with dedicated audio equipment and a broadband Internet connection; unification of hardware reduces installation problems and allows for the creation of config scripts;
    \item performers who cannot use (5) may participate in rehearsals passively (watch / listen only) through a one-way gateway to popular communicators;
    \item remote assistance must be available for all performers at all times, according to the users technical skills; it is worthwhile to test the connection in groups to make sure that the configuration is correct and to fix problems in advance
\end{enumerate}

\vspace{-3mm}

\vspace{-5mm}

\section{Summary}

This paper outlines three proposed Networked Music Performance architectures dedicated to chamber musical ensembles with a conductor, which were defined, deployed and examined in practice. In contrast to other NMP systems and publications, proposed architectures combine uncommon and conflicting requirements of medium-sized choirs, like intra-section feedback, conductor's real-time video, ease of deploy and more. Subsequent architecture defisn iterations were assessed against their usefulness. The conclusive architecture is described along with listed guidelines for its implementation in section \ref{sec:proposed-architecture}.

The Zoom@Home architecture was easy to use; however, it did not allow for rehearsals similar in quality to in-person rehearsals. The Jamulus@Home architecture, implementing software and hardware dedicated to NMP, allowed for actual remote real-time music performance. However, it required much effort from performers, the conductor, and extensive technical assistance. The final Jamulus@University architecture enabled visual input from the conductor and reduced delays between performers, effectively increasing the overall quality of rehearsals, and allowing the choir to effectively work on the new repertoire.

This work proves that it is possible and useful to implement NMP platforms for said musical ensembles using existing affordable tools and equipment. However, overall quality of rehearsals needs to be improved. Work needs to be done to decrease network and audio latencies by incorporating low-latency drivers into operating systems. A lot of effort must be put to lower the entry threshold for NMP tools: automatic configuration and volume control, intuitive user interfaces, integration of video and audio signals, remote assistance helpers. Enabling technologies like peer-to-peer communication, VST plugins, ambisonics and virtual reality techniques would improve musicians' experience even more, allowing to create better NMP platforms in the future.

\label{sec:bib}
\bibliographystyle{splncs04}
\footnotesize
\bibliography{bibliography2}

\begin{thebibliography}{10}
\providecommand{\url}[1]{\texttt{#1}}
\providecommand{\urlprefix}{URL }
\providecommand{\doi}[1]{https://doi.org/#1}

\bibitem{bartlette2006effect}
Bartlette, C., Headlam, D., Bocko, M., Velikic, G.: Effect of network latency
  on interactive musical performance. Music Perception  \textbf{24}(1),  49--62
  (2006)

\bibitem{bouillot2007njam}
Bouillot, N.: {nJam} user experiments: Enabling remote musical interaction from
  milliseconds to seconds. In: Proceedings of the 7th international Conference
  on New interfaces For Musical Expression. pp. 142--147 (2007)

\bibitem{bouillot2009challenges}
Bouillot, N., Cooperstock, J.R.: Challenges and performance of high-fidelity
  audio streaming for interactive performances. In: NIME. pp. 135--140.
  Citeseer (2009)

\bibitem{caceres2008edge}
C{\'a}ceres, J.P., Hamilton, R., Iyer, D., Chafe, C., Wang, G.: To the edge
  with {China}: Explorations in network performance. In: ARTECH 2008:
  Proceedings of the 4th International Conference on Digital Arts. pp. 61--66
  (2008)

\bibitem{chafe2009tapping}
Chafe, C.: Tapping into the {Internet} as an acoustical/musical medium.
  Contemporary Music Review  \textbf{28}(4-5),  413--420 (2009)

\bibitem{chafe2010effect}
Chafe, C., Caceres, J.P., Gurevich, M.: Effect of temporal separation on
  synchronization in rhythmic performance. Perception  \textbf{39}(7),
  982--992 (2010)

\bibitem{chafe2000simplified}
Chafe, C., Wilson, S., Leistikow, R., Chisholm, D., Scavone, G.: A simplified
  approach to high quality music and sound over {IP}. In: Proceedings of the
  COST G-6 Conference on Digital Audio Effects (DAFX-00). pp. 159--164.
  Citeseer (2000)

\bibitem{correia2020evaluating}
Correia, A.P., Liu, C., Xu, F.: Evaluating videoconferencing systems for the
  quality of the educational experience. Distance Education  \textbf{41}(4),
  429--452 (2020). \doi{10.1080/01587919.2020.1821607}

\bibitem{Dessen2020nmpintro}
Dessen, M.: Networked music performance: An introduction for musicians and
  educators. \url{https://ujeb.se/mrEpS5} (2020), [Online; accessed
  15-May-2022]

\bibitem{faadhilah2022comparison}
Faadhilah, A.F., Elfitri, I.: Comparison of audio quality of teleconferencing
  applications using subjective test. In: Audio Engineering Society Convention
  152. Audio Engineering Society (2022)

\bibitem{fischercase}
Fischer, V.: Case study: Performing band rehearsals on the {Internet} with
  {Jamulus}

\bibitem{gu2005network}
Gu, X., Dick, M., Kurtisi, Z., Noyer, U., Wolf, L.: Network-centric music
  performance: Practice and experiments. IEEE Communications  \textbf{43}(6),
  86--93 (2005)

\bibitem{hossfeld2008analysis}
Ho{\ss}feld, T., Binzenh{\"o}fer, A.: Analysis of {Skype} {VoIP} traffic in
  {UMTS}: End-to-end {QoS} and {QoE} measurements. Computer Networks
  \textbf{52}(3),  650--666 (2008)

\bibitem{king2004collaboration}
King, E.C.: Collaboration and the study of ensemble rehearsal. In: Eighth
  International Conference on Music Perception and Cognition (ICMPC8) (2004)

\bibitem{mall2021low}
Mall, P., Kilian, J.: Low-latency online music tool in practice  (2021)

\bibitem{matulin2021comparison}
Matulin, M., Mrvelj, {\v{S}}., Abramovi{\'{c}}, B., {\v{S}}o{\v{s}}tari{\'{c}},
  T., {\v{C}}ejvan, M.: User quality of experience comparison between {Skype},
  {Microsoft} {Teams} and {Zoom} videoconferencing tools. In: Future Access
  Enablers for Ubiquitous and Intelligent Infrastructures. pp. 299--307.
  Springer International Publishing, Cham (2021)

\bibitem{onderdijk2021impact}
Onderdijk, K.E., Acar, F., Van~Dyck, E.: Impact of lockdown measures on joint
  music making: playing online and physically together. Frontiers in psychology
   \textbf{12}, ~1364 (2021)

\bibitem{pennill2019ensembles}
Pennill, N.: Ensembles working towards performance: Emerging coordination and
  interactions in self-organised groups. Ph.D. thesis, University of Sheffield
  (2019)

\bibitem{rottondi2016overview}
Rottondi, C., Chafe, C., Allocchio, C., Sarti, A.: An overview on networked
  music performance technologies. IEEE Access  \textbf{4},  8823--8843 (2016)

\bibitem{volpe2016measuring}
Volpe, G., D'Ausilio, A., Badino, L., Camurri, A., Fadiga, L.: Measuring social
  interaction in music ensembles. Philosophical Transactions of the Royal
  Society B: Biological Sciences  \textbf{371}(1693),  20150377 (2016)

\bibitem{weinberg2005interconnected}
Weinberg, G.: Interconnected musical networks: Toward a theoretical framework.
  Computer Music Journal  \textbf{29}(2),  23--39 (2005)

\end{thebibliography}

\end{document}